\documentclass{elsart}

\usepackage{amssymb}
\usepackage{graphicx}

\begin{document}

\begin{frontmatter}

\title{Characteristics of the $21/2^+$ isomer in $^{93}$Mo:
toward the possibility of enhanced nuclear isomer decay}

\author[a1,a2]{Munetake~Hasegawa,}
\author[a1,a2]{Yang~Sun,}\footnote{Corresponding author at Shanghai Jiao Tong University: sunyang@sjtu.edu.cn}
\author[a3]{Shigeru~Tazaki,}
\author[a4]{Kazunari~Kaneko,}
\author[a5]{Takahiro~Mizusaki}

\address[a1]{Institute of Modern Physics, Chinese Academy of Sciences, Lanzhou 730000, P. R. China}
\address[a2]{Department of Physics, Shanghai Jiao Tong University, Shanghai 200240, P. R. China}
\address[a3]{Department of Applied Physics, Fukuoka University, Fukuoka 814-0180, Japan}
\address[a4]{Department of Physics, Kyushu Sangyo University, Fukuoka 813-8503, Japan}
\address[a5]{Institute of Natural Sciences, Senshu University, Tokyo 101-8425, Japan}


\begin{abstract}

To discuss whether an enhanced isomer decay is a preferred process
in a plasma environment it is required to know the structure of the
isomer as well as the nearby states. The spin-$21/2$, 6.85-hour
high-spin isomer in $^{93}$Mo is investigated within a shell model
which well describes nuclei in this mass region. By using the
obtained wave-functions which correctly reproduce the observed
$B(E2)$, $B(E4)$, and $B(M1)$ transitions, characteristics of the
isomer are shown in comparison with the isomeric states in
neighboring nuclei. Calculations suggest that these high-spin
isomers are formed with almost pure single-particle-like
configurations. The $^{93}$Mo $21/2^+$ isomer has the predominant
configuration $\pi (g_{9/2})^2_{8} \otimes \nu d_{5/2}$ lying below
the $15/2^+$, $17/2^+$, and $19/2^+$ states due to neutron-proton
interaction, which is the physical origin of its long lifetime. The
key $E2$ transition that connects the $21/2^+$ isomer to the upper
$17/2^+$ level is predicted to be substantial (3.5 W.u), and
therefore there is a real prospect for observing induced isomer
deexcitation.

\end{abstract}

\begin{keyword}
High spin isomer \sep Enhanced isomer decay \sep Nuclear shell model

\PACS 23.35.+g, 21.60.Cs, 23.20.Lv, 27.60.+j

\end{keyword}
\end{frontmatter}


Long-lived nuclear isomeric states have been the focus of recent
discussions \cite{Walker,Carroll}. Nuclear isomeric states may play
important roles in nucleosynthesis in stars \cite{Ani}. For the
$^{93}$Mo $21/2^+$ isomer ($E_x =$ 2.425 MeV, $\tau=$ 6.85 h), the
lifetime variation of nuclear levels in a plasma environment
\cite{Gosselin} and possible isomeric triggering via nuclear
excitation by electron capture (NEEC) \cite{Gosselin04,Palffy} have
been suggested. In a hot dense plasma, either laser heated or of
astrophysical sites, nuclei in an isomeric state may have a decay
rate different from the laboratory value because indirect decay
channels may be opened if there exist nuclear levels lying above the
isomeric level that may be excited from the isomeric state, and then
decay down to the ground state.

It has been found experimentally \cite{Fukuchi} that $^{93}$Mo and
the other odd-mass nuclei around it systematically have high-spin
isomers (the $21/2^+$ isomer and others). It is qualitatively
understood that high-spin isomers in the nuclei near shell closure
appear when neutron number and/or proton number is odd, namely, when
the number of valence neutrons outside the $N$ = 50 closed shell is
one or three and/or proton number is $Z$ = 40 plus one or three.
Although in the present case the proton number $Z$ = 40 is a
semi-magic number, a few protons near $Z$ = 40 seem to play a
leading role in the formation of isomeric structure. Thus a few
valence protons and neutrons outside a magic or semi-magic shell
participate in the construction of the isomeric states in and near
$^{93}$Mo. The purpose of the present Letter is to show
characteristics of the isomeric states, especially for the $21/2^+$
isomer in $^{93}$Mo. Understanding this isomer and its possible
decay channels is important for the suggested enhanced isomer decay
\cite{Gosselin,Gosselin04,Palffy}. For some key transition
probabilities, especially the one that links the $21/2^+$ isomer to
the upper-lying $17/2^+$ state, there has been so far no
experimental measurement or theoretical calculation available. The
discussion of lifetimes of $^{93}$Mo in hot dense plasmas was based
on an assumption for the unknown transition rate \cite{Gosselin}.


To obtain a quantitative understanding of the isomer decay, shell
model calculations that can give detailed information about the
microscopic insight of the states are much desired. Good effective
interactions generally applicable to this mass region are needed.
Ref. \cite{Sieja09} employed an effective interaction derived by
monopole corrections of the realistic $G$ matrix and studied the
low-lying spectra of Zr isotopes.  There was an early shell-model
study \cite{Zhang99} for nuclei with $A=92-98$ including $^{93}$Mo.
This article, however, focused on the discussion of the truncation
scheme of spherical shell model, but did not investigate the
structure and electromagnetic properties of the isomeric state
$21/2^+$ and other high-spin states in $^{93}$Mo. So far, knowledge
of these states that is decisive in the isomer studies in a plasma
environment has not been attained. Electromagnetic transition
properties related to the $21/2^+$ isomer in $^{93}$Mo have not been
investigated. Recently, an extended $P+QQ$ interaction with monopole
corrections \cite{Hase1,Hase2} has been applied to interpret new
experimental data of $^{94}$Mo and $^{95}$Mo \cite{Zhang}. The model
reproduces well the observed level scheme up to quite high spins in
$^{94,95}$Mo. The shell model parameters are determined so as to
consistently reproduce overall energy levels of the $40 \le Z \le
42$ and $50 \le N \le 53$ nuclei. The region covers the nuclei
around $^{93}$Mo which are the targets of the present Letter. This
shell model, therefore, provides an appropriate tool for our
purpose.

The model is outlined as follows; for detail, readers are referred
to Ref. \cite{Zhang}. For a general consideration we take 8 orbits
($f_{5/2}$, $p_{1/2}$, $g_{9/2}$, $d_{5/2}$, $s_{1/2}$, $d_{3/2}$,
$g_{7/2}$, $h_{11/2}$) outside $^{64}$Ge as the valence space, with
fixed single-particle energy (all in MeV)
 $\varepsilon_{f5/2} = -0.85$,
 $\varepsilon_{p1/2} = 0.0$,  $\varepsilon_{g9/2} = 1.90$,
 $\varepsilon_{d5/2} = 4.20$, $\varepsilon_{s1/2} = 6.06$,
 $\varepsilon_{d3/2} = 6.63$, $\varepsilon_{g7/2} = 6.70$,
 $\varepsilon_{h11/2} = 7.90$.
In the present calculation we restrict the model space so that
valence protons act in the orbits ($p_{1/2}$, $g_{9/2}$, $d_{5/2}$)
and valence neutrons in ($d_{5/2}$, $s_{1/2}$, $d_{3/2}$, $g_{7/2}$,
$h_{11/2}$). One must notice that the proton-hole orbit $f_{5/2}$
and the neutron-hole orbits $(f_{5/2},p_{1/2}, g_{9/2})$ change the
{\it particle} energies, and therefore have effects on calculated
results.

The shell model code NuShellX \cite{Rae} is employed. For the
interactions, mass ($A$) dependent force strengths for the $J=0$ and
$J=2$ pairing forces, quadrupole-quadrupole (QQ) force,
octupole-octupole (OO) force are respectively fixed as follows:
 $g_0 = 25/A$, $g_2 = 260/A^{5/3}$,
 $\chi_2 = 300/A^{5/3}$, $\chi_3 = 200/A^6$ for the pp terms;
 $g_0 = 20/A$, $g_2 = 260/A^{5/3}$,
 $\chi_2 = 200/A^{5/3}$, $\chi_3 = 200/A^6$ for the nn and np terms.
Here, we use the same force strengths for the nn and np interactions
in order to reduce the number of parameters. Moreover, we add two
monopole corrections to the proton interactions,
 $\Delta k^{T=1}(p_{1/2}^\pi,p_{1/2}^\pi) = -0.45$ and
 $\Delta k^{T=1}(g_{9/2}^\pi,g_{9/2}^\pi) = -0.25$ in MeV.


\begin{figure}[b]
\includegraphics[width=10cm,height=13cm]{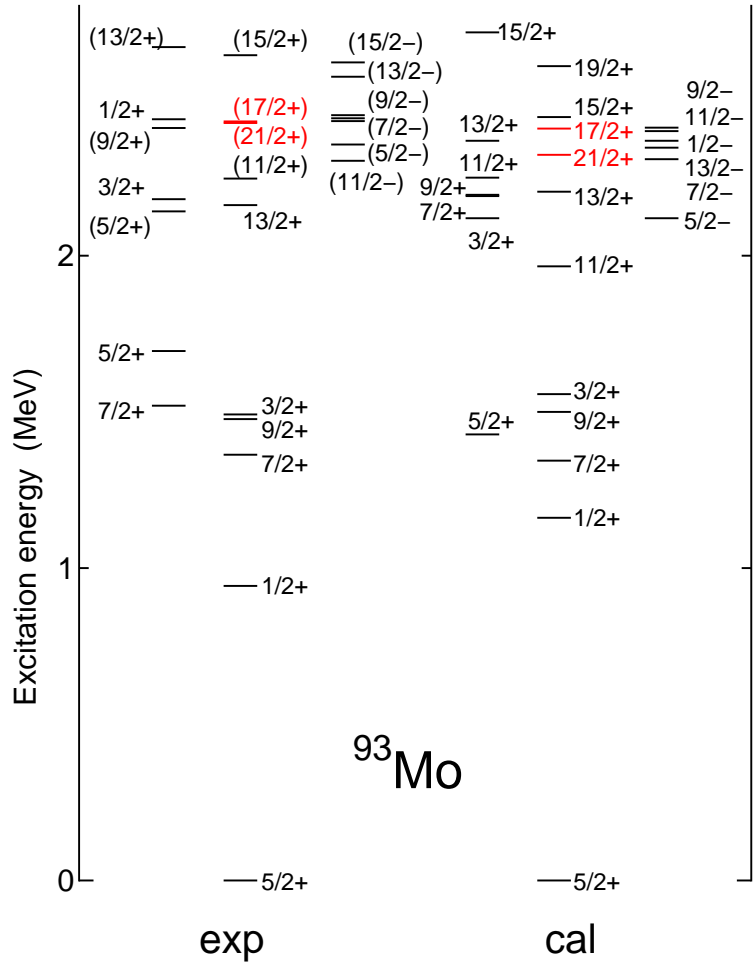}
  \caption{(Color online) Calculated energy levels of $^{93}$Mo,
           which are compared with experimental data taken from
           Ref. \protect\cite{ENSDF}.
           }
  \label{fig1}
\end{figure}

Calculated energy levels for $^{93}$Mo are shown in Fig. \ref{fig1},
which are compared with the experimental data taken from the
evaluated nuclear structure data file \cite{ENSDF}. It can be seen
that the present calculation not only reproduces well the energy
levels of the yrast states, but also lays the non-yrast positive-
and negative-parity states at correct energies. The model describes
the essential condition for the $21/2_1^+$ state to be an isomer
because the lower spin $15/2_1^+$, $17/2_1^+$, and $19/2_1^+$ states
all lie above the $21/2_1^+$ state, and therefore, the maximum spin
below the $21/2_1^+$ state is $J^\pi =13/2^+$. Multiplicities in
electromagnetic decay of the $21/2_1^+$ state are therefore $J \ge
4$; namely, the most probable decay of the $21/2_1^+$ state is an
$E4$ transition to the $13/2_1^+$ state. The calculation also
correctly predicts that the nearest positive-parity level above the
$21/2_1^+$ state is the $17/2_1^+$ state, although the calculated
energy difference is a bit larger than the observed one (5 keV). The
experimental level at 2.247 MeV between $13/2_1^+$ and $21/2_1^+$
was temporarily assigned as $11/2^+$ while our model predicts the
second $11/2^+$ state at the corresponding energy. However, if the
state observed at 2.247 MeV would be the yrast $11/2_1^+$ state, our
model fails to reproduce the inverse order of the $11/2^+$ and
$13/2^+$ levels. Except for this uncertainty, the successful
calculation allows us to discuss the structure of $^{93}$Mo.

In Table \ref{table1}, we show the main components of wave-functions
for the yrast states in $^{93}$Mo with spin $J^+ \le 21/2^+$.
Leading configurations in these states are given by $\pi
(p_{1/2},g_{9/2},d_{5/2})^4 [J_p^+] \otimes \nu d_{5/2}$, in which
we specify configurations of four protons by spin-parity $J_p^+$.
Table \ref{table1} clearly shows that the structure change in the
positive-parity yrast states is caused by the change in proton
configurations. The $21/2^+$ isomer is almost in the configuration
$\pi [8^+] \otimes \nu d_{5/2}$, 92\% of which is $\pi
(g_{9/2})^2_{J=8} \otimes \nu d_{5/2}$. The main configuration of
the $17/2_1^+$ state (lifetime 3.53 ns) has 59.1\% in $\pi
(g_{9/2})^2_{J=6} \otimes \nu d_{5/2}$ and 26.7\% in $\pi
(g_{9/2})^2_{J=8} \otimes \nu d_{5/2}$. Thus the main structure of
these two isomeric states is of a rather simple single-particle
configuration $\pi (g_{9/2})^2 \otimes \nu d_{5/2}$. Specifically,
the $21/2^+$ isomer is formed with the fully aligned spin of two
$g_{9/2}$ protons and one neutron, which characterizes a high-spin
isomer.

\begin{table}[b]
\caption{Structure of the yrast states in $^{93}$Mo.
         Only the leading configurations
         $\pi (p_{1/2},g_{9/2},d_{5/2})^4 [J_p^+]$ $\otimes \nu d_{5/2}$
         are shown, with $J_p^+$ denoting the spin of four valence protons.
         The squared amplitudes (in percent) for each $J_p^+$ are given.}
\begin{tabular}{|c|ccccc|}   \hline
 yrast & \multicolumn{5}{c|}{spin of four valence protons} \\
 state & $0^+$ & $2^+$ & $4^+$ & $6^+$ & $8^+$  \\ \hline

$5/2^+$   & 92.5 &  5.9 &      &      &       \\
$1/2^+$   & 51.4 & 47.1 &      &      &       \\
$7/2^+$   &      & 89.5 &  8.0 &      &       \\
$9/2^+$   &      & 88.2 &  8.6 &  1.1 &       \\
$3/2^+$   &  9.2 & 88.0 &  1.4 &      &       \\
$11/2^+$  &      &      &  9.4 & 12.8 & 74.8  \\
$13/2^+$  &      &      & 34.8 & 23.1 & 39.3  \\
$21/2^+$  &      &      &      &      & 99.5  \\
$17/2^+$  &      &      &      & 67.1 & 29.2  \\
$15/2^+$  &      &      &      & 40.6 & 56.8  \\
$19/2^+$  &      &      &      &      & 98.8  \\ \hline
\end{tabular}
\label{table1}
\end{table}

Lifetimes of the isomeric states depend on the reduced
electromagnetic transition probabilities and energy spacings from
the lower states. The $21/2^+$ isomer decays to the $13/2_1^+$ state
by an $E4$ transition with the largest probability. The $17/2_1^+$
state can decay to the $21/2_1^+$ state by an $E2$ transition but
the energy difference (5 keV) is too small, and hence it decays
mainly to the $13/2_1^+$ state by an $E2$ transition. On the other
hand, the $21/2^+$ and $17/2^+$ isomeric states cannot take a course
of $M1$ decay because the $19/2_1^+$ ($15/2_1^+$) level lies above
the $21/2_1^+$ ($17/2_1^+$) level.


Around $^{93}$Mo, the nuclei with odd-number neutrons $^{91}$Zr,
$^{95}$Ru, and $^{97}$Ru have also a $21/2_1^+$ isomer (with
lifetimes 4.35 $\mu$s, 10.05 ns, and 7.8 ns, respectively).
$^{95}$Ru has another isomeric state $17/2_1^+$ with lifetime 3.05
ns. We have carried out shell model calculations for these nuclei
with the same calculation conditions. The results show that the
isomeric states $21/2_1^+$ and $17/2_1^+$ in these nuclei all have
the single-particle configuration $\pi (g_{9/2})^2 \otimes \nu
d_{5/2}$ as the main component, as those in $^{93}$Mo. However, the
most significant difference of these nuclei from $^{93}$Mo is that
the $17/2_1^+$ level lies below the $21/2_1^+$ level. Therefore,
lifetime of the $21/2_1^+$ state in $^{91}$Zr, $^{95}$Ru, and
$^{97}$Ru is much shorter than that in $^{93}$Mo.


It has been experimentally known that the nuclei with odd-number
protons, $^{91}$Nb and $^{93}$Tc, have also a $21/2^+$ isomer, with
lifetime 0.92 ns and 1.72 ns, respectively. $^{93}$Tc is the isotone
of $^{93}$Mo which can be created from $^{93}$Mo by changing a
neutron into a proton. In another odd-proton nucleus, $^{93}$Nb,
$21/2^+$ isomer has not been observed while calculations predicted
its existence \cite{Hori}. For a deeper understanding of the
$21/2^+$ isomer, it is interesting to compare the one in $^{93}$Mo
with that in $^{93}$Tc, which is now calculated with the same shell
model.

\begin{figure}[b]
\includegraphics[width=10cm,height=13cm]{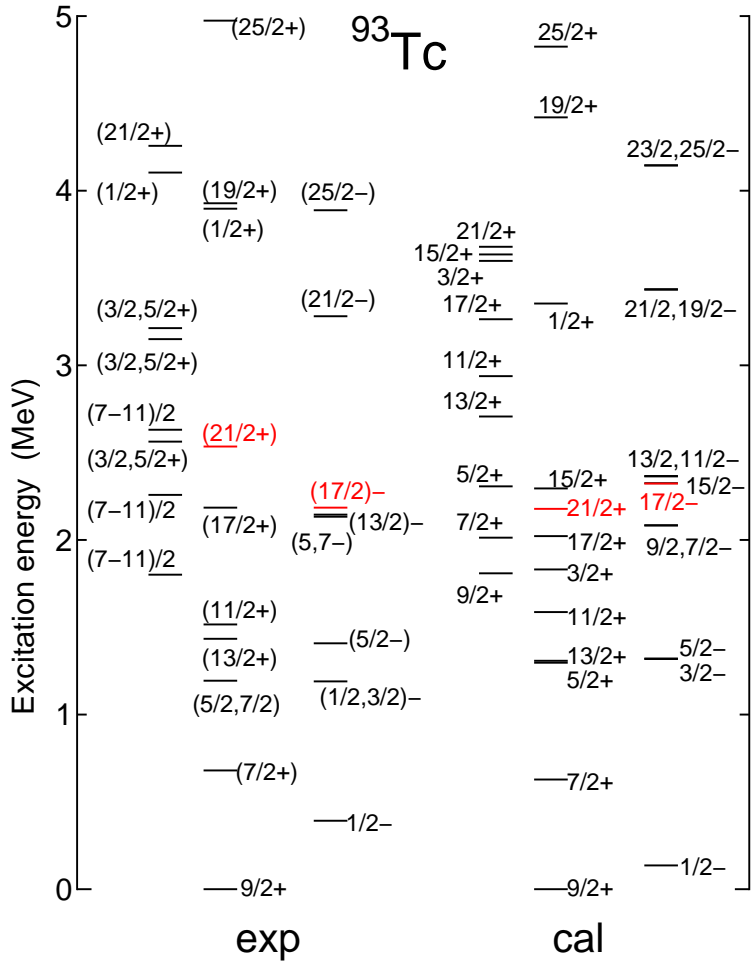}
  \caption{(Color online) Calculated energy levels of $^{93}$Tc,
           which are compared with experimental data taken from
           Ref. \protect\cite{ENSDF}.}
  \label{fig2}
\end{figure}

Figure \ref{fig2} shows experimental and calculated energy levels
for $^{93}$Tc. One sees that the model describes well the observed
levels also for this odd-proton nucleus. It reproduces the order of
the positive-parity yrast states and also other positive- and
negative-parity states. The theory lays correctly the $17/2_1^+$
level below the $21/2_1^+$ state but the $19/2_1^+$ (and $15/2_1^+$)
level above the $21/2_1^+$ state. This condition prohibits the decay
of the $21/2_1^+$ state to the $19/2_1^+$ state by $E2$ or $M1$
transitions but allows only a decay to the $17/2_1^+$ state by $E2$
transition. We can thus understand some retardation in the decay of
the $21/2_1^+$ state in $^{93}$Tc.

What structure does the isomeric state $21/2^+$ have in $^{93}$Tc?
In the present model, the $^{93}$Tc nucleus has no valence neutron,
and therefore the states are generally described by the
configuration with five protons $\pi (p_{1/2},g_{9/2},d_{5/2})^5
[J^+]$. Calculations show that except for the $19/2^+$ and $1/2^+$
states {\footnote{ Because the configuration $\pi (g_{9/2})^3[J^+] $
has no state when $J$=19/2 and $1/2$, the $19/2^+$ and $1/2^+$
states must have the main configuration $\pi (g_{9/2})^5$.}} the
main component of the yrast states up to $21/2^+$ is of the
configuration $\pi (p_{1/2})^2(g_{9/2})^3 [J^+]$, the percentage of
which is tabulated in Table \ref{table3}. The isomeric state
$21/2^+$ has the single-particle configuration $\pi (g_{9/2})^3
[21/2^+]$ as the main structure. The spin $21/2$ is attained by
alignment of three protons in the $g_{9/2}$ orbit
(9/2+7/2+5/2=21/2). The calculation for $^{91}$Nb gives the same
order of energy levels with respect to spin as in $^{93}$Tc. Thus we
have understood that the isomeric state $21/2^+$ in the proton-odd
nuclei is given by the predominant configuration $\pi (g_{9/2})^3
[21/2^+]$.

\begin{table}
\caption{Structure of the yrast states with spin $1/2^+ < J^\pi \le 21/2^+$
         in $^{93}$Tc. The percentage of the main configuration
         $\pi (p_{1/2})^2(g_{9/2})^3 [J^+]$ for each yrast state is shown.}
\begin{tabular}{|c|c||c|c||c|c||c|c|}   \hline
 state &    & state &    & state &    & state &     \\ \hline
$3/2^+$   & 89.6 &  $9/2^+$ & 82.1 & $15/2^+$ & 91.3 & $21/2^+$ & 93.1  \\
$5/2^+$   & 84.2 & $11/2^+$ & 92.8 & $17/2^+$ & 92.7 &          &       \\
$7/2^+$   & 88.8 & $13/2^+$ & 89.2 & $19/2^+$ &  -   &          &       \\ \hline
\end{tabular}
\label{table3}
\end{table}

The difference in the level scheme of $^{93}$Mo and $^{93}$Tc,
especially the inverse order of $21/2_1^+$ and $17/2_1^+$ between
the two isotones, is considered to come from the different effects
between the np interaction in the coupling $\pi (g_{9/2})^2_{8}
\otimes \nu d_{5/2}$ and the pp interaction in $\pi (g_{9/2})^2_{8}
\otimes \pi g_{9/2}$. The long lifetime of the $^{93}$Mo $21/2^+$
isomer is thus interpreted as due to the np coupling.

The $^{93}$Tc nucleus has another isomeric state $17/2^-$ with
lifetime 10.2 $\mu$s. Our calculation lays the $17/2_1^-$ state
below the $13/2_1^-$ state, in disagreement with the observed order.
However, as the difference is small in energy, the obtained
wave-function can still be reliable. The calculated $17/2_1^-$ state
has 96.5\% in the configuration $\pi p_{1/2}(g_{9/2})^4 [17/2^-]$.
The $^{91}$Nb nucleus has a similar level scheme for negative-parity
states and an isomeric state with spin $17/2^-$ (lifetime 3.76
$\mu$s). Our model reproduces equally well the observed energy
levels including the isomeric state $17/2_1^-$. The isomeric state
$17/2_1^-$ in both $^{93}$Tc and $^{91}$Nb has the spin-aligned
three-proton configuration $\pi p_{1/2} \otimes \pi (g_{9/2})^2_{8+}
[17/2^-]$. The results altogether suggest that the isomeric states
in odd-mass nuclei around $^{93}$Mo are all characterized by a
spin-aligned configuration in which a single neutron or proton
couples with a fully aligned proton pair in the $g_{9/2}$ orbit
($(g_{9/2})^2_{8+}$).

The successful calculation for energy levels encourages us to
predict electromagnetic transition probabilities for $^{93}$Mo,
which are relevant to the discussions about the lifetime variation
of nuclear isomers in a plasma environment \cite{Gosselin} and the
NEEC effect \cite{Gosselin04,Palffy}. In Table \ref{table2}, we show
calculated reduced electromagnetic transition probabilities, in
which the standard effective charges $e_p=1.5e$ and $e_n=0.5e$ for
electric transitions are used. For magnetic transitions we employ
the quenched spin g-factors $g^s_{p}=3.18$ and $g^s_{n}=-2.18$,
which were used in Ref. \cite{Lisetskiy} to explain the observed
$B(M1)$ values in $^{94}$Mo. As seen in Table \ref{table2}, all the
calculated values of $B(E4)$ or $B(E2)$ as well as $B(M1)$ agree
well with the known data. (The experimental value 417 W.u.
\cite{ENSDF} for $B(E2;13/2^+ \rightarrow 9/2^+)$ is apparently too
large, though.) In addition, many unknown transitions are predicted.

\begin{table}
\caption{Experimental and calculated reduced electromagnetic
transition
         probabilities in $^{93}$Mo related to the decay of the isomeric states
         $21/2^+$ and $17/2^+$.
         The $B(E4)$, $B(E2)$, and $B(M1)$ values are shown in W.u.}
\begin{tabular}{|c|cc|cc|}   \hline
       & \multicolumn{2}{c|}{$B(E4)$ or $B(E2)$}
       & \multicolumn{2}{c|}{$B(M1)$}                 \\
 $J_i \rightarrow J_f$  & exp. & cal.  & exp. & cal.  \\ \hline
$21/2^+ \rightarrow 13/2^+$ & 1.431 (24) &  1.9  &            &       \\
$17/2^+ \rightarrow 21/2^+$ &            &  3.5  &            &       \\
$17/2^+ \rightarrow 13/2^+$ & 4.48 (23)  &  4.0  &            &
\\ \hline
$19/2^+ \rightarrow 15/2^+$ &            &  2.5  &            &       \\
$19/2^+ \rightarrow 17/2^+$ &            &  0.01 &            & 0.34  \\
$19/2^+ \rightarrow 21/2^+$ &            &  0.03 &            & 0.84  \\
$15/2^+ \rightarrow 17/2^+$ &            &  0.02 &            & 1.07  \\
$15/2^+ \rightarrow 11/2_1^+$ &          &  2.2  &            &       \\
$15/2^+ \rightarrow 13/2^+$ &            &  0.01 &            & 0.85
\\ \hline
$13/2^+ \rightarrow  9/2^+$ & 417 (93)   &  7.2  &            &       \\
$ 9/2^+ \rightarrow  7/2^+$ &            &  6.5  & 0.38 (13)  & 0.47  \\
$ 9/2^+ \rightarrow  5/2^+$ & 12 (4)     & 12.9  &            &       \\
$ 7/2^+ \rightarrow  5/2^+$ & 8.7 (22)   & 12.6  & 0.068 (6)  &
0.037 \\ \hline
$ 3/2^+ \rightarrow  1/2^+$ &            &  0.51 &            & 0.31  \\
$ 1/2^+ \rightarrow  5/2^+$ & $76^{+75}_{-56}$ & 25  &        &       \\
$ 3/2^+ \rightarrow  5/2^+$ &            &  4.9  &            & 0.22
\\ \hline
\end{tabular}
\label{table2}
\end{table}

The $B(E2)$ values related to the $21/2^+$ and $17/2^+$ states
indicate that these isomeric states are not collective, as expected
from their fully-aligned configuration $\pi (g_{9/2})^2 \otimes \nu
d_{5/2}$. On the other hand, Table \ref{table2} indicates a larger
collectivity for the low-lying low-spin states. For example, the
calculation gives the value $B(E2;1/2^+ \rightarrow 5/2^+)=25$ W.u.
for the observed value $76^{+75}_{-56}$ W.u, and therefore, the
ground state $5/2^+$ and the first excited state $1/2^+$ are more
collective. We stress that our model not only describes the isomeric
states $21/2^+$ and $17/2^+$ at high energy but also those low-lying
states.

Enhanced decay of the $21/2^+$ isomer in $^{93}$Mo, as suggested in
Refs. \cite{Gosselin,Gosselin04,Palffy}, involves an E2 transition
to the upper-lying state $17/2^+$. It is thus crucial to known the
E2 transition probability of the $21/2^+$ isomer to the upper-lying
state. This transition has not been known experimentally. In the
discussion of lifetimes of $^{93}$Mo in hot dense plasmas
\cite{Gosselin}, the experimental values of $B(E4;21/2_1^+
\rightarrow 13/2_1^+)$ and $B(E2;17/2_1^+ \rightarrow 13/2_1^+)$
were used but the unknown $B(E2;17/2_1^+ \rightarrow 21/2_1^+)$
value was {\it assumed} to be the same as $B(E2;17/2_1^+ \rightarrow
13/2_1^+)=4.48$ W.u. As shown in Table \ref{table2}, our model
predicts a value 3.5 W.u. for the transition $B(E2;17/2_1^+
\rightarrow 21/2_1^+)$, which is close to the assumed value in Ref
\cite{Gosselin}. Although the states are not collective, the
predicted $B(E2)$ for the 21/2-to-17/2 transition is quite
substantial, and there is therefore a real prospect for observing
induced isomer deexcitation suggested by Gosselin {\it et al.}
\cite{Gosselin}.


In conclusion, using a modern shell model, we have performed
microscopic shell model calculations for the $^{93}$Mo $21/2^+$
isomer and compared it with the $21/2^+$ and other spin states in
the odd-mass nuclei around $^{93}$Mo. The calculations have
confirmed the conclusion from an early simpler calculation that
these isomeric states have rather simple single-particle
configurations of 2-proton+1-neutron or 3 protons as the main
structure, which are characterized by fully aligned spins. The
predominant configuration of the $^{93}$Mo $21/2^+$ isomer is $\pi
(g_{9/2})^2_{J=8} \otimes \nu d_{5/2}$. We have calculated
electromagnetic transition probabilities between the $21/2^+$ isomer
and the neighboring states using the detailed shell-model
wavefunctions, thus providing the important structure information
that has been missing so far in the isomer decay studies. It has
been found that the small mixtures of complicated configurations
determine the details of the transition probabilities. The long
lifetime of the $^{93}$Mo isomer is attributed to the fact that the
$21/2^+$ state lies below the $15/2_1^+$, $17/2_1^+$, and $19/2_1^+$
states and hence the major electromagnetic transitions $E2$ and $M1$
are prohibited. The predicted E2 transition probability of the
isomer to the upper-lying $17/2_1^+$ provides a rather positive
support to the proposed enhancement of isomer decay in a plasma
environment.

Useful discussions with Dr. Phil Walker and Dr. G. Gosselin are
acknowledged. One of us (MH) thanks colleagues of Department of
Physics for the hospitality extended to him when he visited Shanghai
Jiao Tong University (SJTU), and he is grateful for support from the
SJTU-INS Research Project for Visiting Scholars. Research at SJTU is
supported by the Shanghai Pu-Jiang grant, the National Natural
Science Foundation of China under contract No. 10875077, and the
Chinese Major State Basic Research Development Program through grant
No. 2007CB815005.



\end{document}